\documentclass[
    ,final            
  ]
  {aipproc}

\layoutstyle{6x9}
\usepackage{epsf,graphics,amssymb,amsfonts,amsmath,array}

\def\lQ{\Lambda_{\rm QCD}}

\def\als{\alpha_{\rm s}}
\newcommand{\MS}{\overline{\rm MS}}
\def\siml{{\ \lower-1.2pt\vbox{\hbox{\rlap{$<$}\lower6pt\vbox{\hbox{$\sim$}}}}\ }} 
\def\simg{{\ \lower-1.2pt\vbox{\hbox{\rlap{$>$}\lower6pt\vbox{\hbox{$\sim$}}}}\ }}

\def\vbfD{{\ \lower-8pt\vbox{\hbox{\rlap{$\!\leftrightarrow$}\lower8pt\vbox{\hbox{$\!\bf D$}}}}\ }} 
\def\dsl{\,\raise.15ex\hbox{/}\mkern-13.5mu D}

\newcommand{\nn}{\nonumber}
\newcommand{\be}{\begin{equation}} 
\newcommand{\ee}{\end{equation}}
\newcommand{\bea}{\begin{eqnarray}} 
\newcommand{\eea}{\end{eqnarray}}
\newcommand{\beq}{\begin{equation}}
\newcommand{\eeq}{\end{equation}}
\newcommand{\bqa}{\begin{eqnarray}}
\newcommand{\eqa}{\end{eqnarray}}

\begin{document}

\title{The QCD potential}

\classification{12.38.-t, 12.38.Bx, 12.38.Gc, 12.39.Hg}
\keywords      {effective field theories, pNRQCD, potential, perturbative QCD, lattice}

\author{Antonio Vairo}{
  address={Dipartimento di Fisica dell'Universit\`a di Milano and INFN, via
  Celoria 16, 20133 Milano, Italy \\
  IFIC, Universitat de Val\`encia-CSIC, Apt. Correus 22085, E-46071 Val\`encia, Spain}}

\begin{abstract}
After reviewing the definition of the heavy quark-antiquark potential  
in pNRQCD, we discuss recent advances in the calculation. 
\end{abstract}

\maketitle

\section{Definition}
The potential between a heavy quark and antiquark has been one of the first 
quantities to be studied in QCD: it is a privileged object for 
exploring the interplay of perturbative and non-perturbative
QCD and the set in of confinement, and it plays a central role 
in quarkonium physics \cite{Brambilla:2004wf}.
Nowadays, the progress of perturbative and lattice calculations requires
an accurate and rigorous definition of the potential in QCD, phenomenological 
and intuitive characterizations being no longer adequate.

So, what is the QCD potential between a quark and antiquark with a large mass $m$? 
One may first answer that the potential is the function $V$ into
the Schr\"odinger equation describing the quark-antiquark bound state $\phi$:
\be
E \, \phi = \left( \frac{p^2}{m}+  V\right)\,\phi, 
\label{schroe1}
\ee
$p$ being the momentum of the quark-antiquark pair in the centre-of-mass 
system and $E$ its binding energy. Clearly, if Eq. (\ref{schroe1}) comes 
from a systematic expansion of QCD, it arises from at least a double expansion 
in $p/m$ or $r m $ ($r$ being the inter-quark distance) and in $E\,r$.
Hence, rather than Eq. (\ref{schroe1}), we may expect that the QCD expansion 
would lead to 
\be
E \, \phi = \left( \frac{p^2}{m} +  V^{(0)}(r) +  \frac{V^{(1)}(r)}{m} + \dots  
\right)\,\phi, 
\label{schroe2}
\ee
where the $\dots$ stand both for terms suppressed in the non-relativistic 
expansion in  $p/m$ or $r m$ and for terms suppressed in  $E\,r$,  
sometimes referred to as retardation effects (an example is the Lamb-shift).
The above double expansion becomes an expansion in the heavy-quark velocity 
$v$ once we note that in a non-relativistic system $1/r \sim p \sim mv$, 
and $E \sim mv^2$, with  $v \ll 1$.

\begin{figure}[ht]
\makebox[-5truecm]{\phantom b}
\put(20,0){\epsfxsize=6truecm \epsfbox{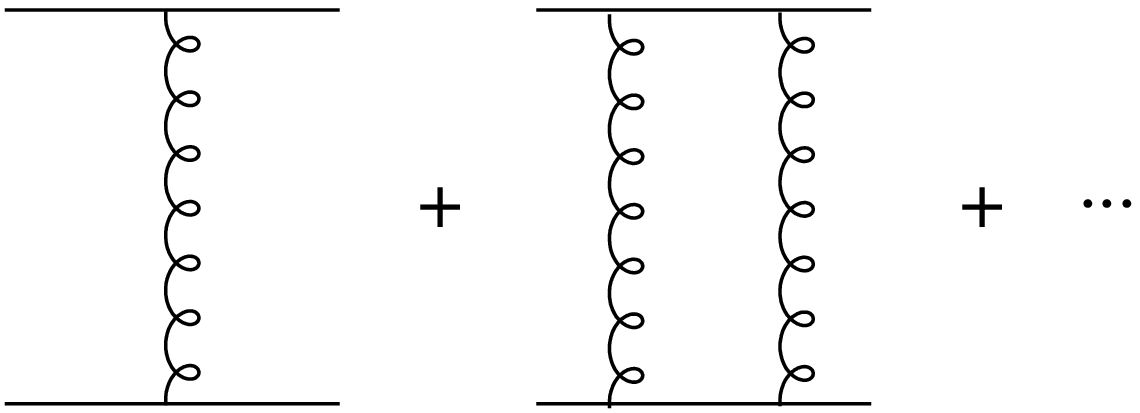}}
\put(200,25){$\displaystyle \approx \frac{1}{E - \left(\frac{p^2}{m} + V\right)}$}
\put(38,-19){$\als\left(1+\als/v + \dots \right)$ }
\caption{Resummed propagator near threshold.
\label{figcoulomb}}
\end{figure}

How do the scales $mv$ and $mv^2$ originate in QCD? Let's consider the case 
of weakly-coupled bound states, i.e. states such that $\lQ$ is smaller  
than any of the scales $m$, $mv$ or $mv^2$. For these states we may 
use perturbation theory. Near threshold, the momenta of the quarks 
are small compared to their masses, so that $p/m \sim v \ll 1$. 
Moreover, for certain sets of graphs, like those in Fig. \ref{figcoulomb}, 
the perturbative expansion breaks down when $\als \sim v$. 
The summation of all $\als/v$ contributions leads
to the appearance of a bound-state pole of order $mv^2\sim m\als^2$ 
in the resummed propagator.  

\begin{figure}[hb]
\makebox[-5truecm]{\phantom b}
\put(20,0){\epsfxsize=7truecm \epsfbox{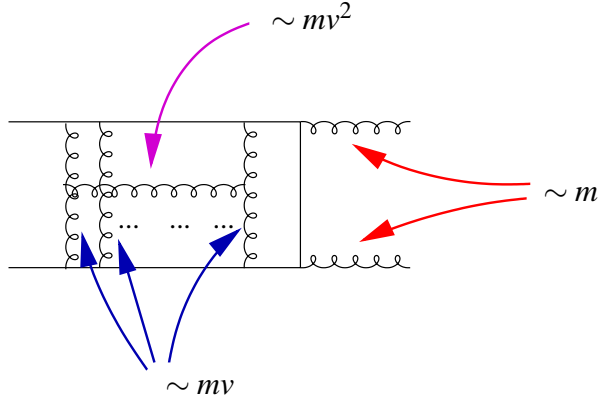}}
\put(80,-8){$\sim mv$}
\put(120,130){$\sim mv^2$}
\put(223,65){$\sim m$}
\caption{Annihilation diagram contributing to the quarkonium decay width.
\label{figentangled}}
\end{figure}

These scales get entangled in a typical amplitude. 
An example is provided by the annihilation diagram of Fig. \ref{figentangled}. 
Assuming that the incoming quarks are near threshold, the different gluons entering 
the diagram are characterized by different scales. The annihilation gluons 
have a typical energy of order $m$, sometimes also called ``hard scale''; 
binding gluons, also called ``soft'',  have the momentum of the incoming quarks, 
which is of order $mv$, and ``ultrasoft'' gluons, sensitive to the intermediate 
bound state, have energies of the order of the binding energy, i.e. $mv^2$.

In order to disentangle the different scales, it is convenient to enforce 
an expansion in the ratios of low-energy scales over large-energy 
scales at the Lagrangian level; this corresponds to substituting QCD 
with low-energy Effective Field Theories (EFTs)
\cite{Brambilla:2004jw}. The ultimate EFT that follows from QCD by integrating out all 
energy scales but $mv^2$ is potential NRQCD (pNRQCD).
The general form of the Lagrangian density of pNRQCD is 
\be
{\cal L} = \int d^3r \; S^\dagger \left( i\partial_0 - 
\frac{p^2}{m} -  V^{(0)}_s(r,\mu) -  \frac{V^{(1)}_s(r,\mu)}{m} + \dots  
\right)S + \hbox{ultrasoft contributions}, 
\label{pNRQCD}
\ee
where $S$ stands for a color-singlet quarkonium field, $\mu$ is the cut-off of the EFT,  
and ``ultrasoft contributions'' include all degrees of freedom which are 
ultrasoft (they may be gluons, or light quarks or other degrees 
of freedom). The ultrasoft contributions are typically suppressed with respect 
to the part of the pNRQCD Lagrangian displayed in Eq. (\ref{pNRQCD}). 
Hence, the equation of motion of the color-singlet 
quarkonium field is exactly Eq. (\ref{schroe2}) and we may identify $ V^{(0)}(r,\mu) + 
V^{(1)}(r,\mu)/m + \dots$ with the heavy-quark potential. 

In summary, EFTs provide the following definition of the potential:
the potential is a Wilson coefficient of the EFT obtained by integrating out 
all degrees of freedom but the ultrasoft ones, it  undergoes renormalization, 
develops a scale dependence and satisfies renormalization
group equations, which eventually allow to resum potentially large logarithms.

\section{The Perturbative Potential}
\begin{figure}[ht]
\makebox[-7cm]{\phantom b}\put(0,0){{\epsfxsize=14truecm\epsffile{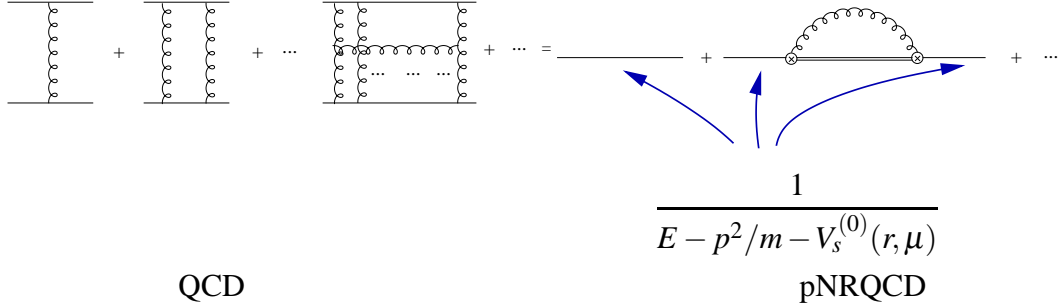}}}
\put(65,-55){QCD}\put(300,-55){pNRQCD}
\put(245,-25){$\displaystyle \frac{1}{E - p^2/m-V_s^{(0)}(r,\mu)}$}
\caption{Matching condition for pNRQCD. On the left-hand side the four-fermion Green's function 
in QCD, on the right-hand side the singlet propagator and the first ultrasoft correction in pNRQCD.
The single continuous line stands for a singlet propagator, the double line for an octet 
propagator and the curly line for a chromoelectric correlator. The coupling of the gauge fields 
with the quarkonium (circle with a cross) is a chromoelectric dipole vertex.
\label{figmatch}}
\end{figure}

We consider the static potential $V^{(0)}_s$. This is obtained by integrating 
out soft gluons from static QCD. Soft gluons are those associated with the scale $1/r$.
At short distances, $1/r \gg \lQ$, soft gluons may be calculated in perturbation
theory. If also the ultrasoft scale, i.e. the potential itself, is larger than $\lQ$, 
then, besides the color-singlet quarkonium field, ultrasoft degrees of freedom 
include ultrasoft gluons and the color-octet quarkonium field.  The matching  
leading QCD to pNRQCD may be done in perturbation theory, see Fig. \ref{figmatch}. 
Sometimes it may be useful to choose the QCD Green's function in  a gauge invariant fashion. 
A popular choice is the static Wilson loop.

The matching fixes the potential and the other Wilson coefficients of the EFT.
The matching condition for the singlet static potential reads
\bea
\lim_{ T\to\infty}\frac{i}{T } \ln  
\put(0,-5){\epsfxsize=1.2cm \epsffile{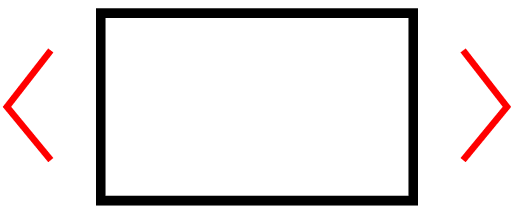}}
\hspace{1.2cm}
&=&  V_{s}^{(0)}(r,\mu) 
\nn\\
&& - i \frac{g^2}{ N_c} V_A^2 \,\frac{r^2}{3} \,
\int_0^\infty \!\!\! dt \, e^{-i t  (V^{(0)}_o-V^{(0)}_s)} \,   
 \langle  {\rm Tr}( {\bf E}(t) \cdot  {\bf E}(0))  \rangle(\mu)  + \dots, 
\label{matchingVs}
\eea
where the box stands for the static Wilson loop of dimension $r\times T$, $V^{(0)}_o$ 
for the static octet potential, $V_A$ for the electric-dipole matching coefficient, 
${\bf E}$ for the chromoelectric field and $N_c=3$ for the number of colors.
The left-hand side of Eq. (\ref{matchingVs}) is known at two loops 
\cite{Peter:1996ig,Peter:1997me,Schroder:1998vy,Kniehl:2001ju}. 
At three loops the static Wilson loop contains a term proportional 
to $\als^4/r \times \ln \als$, which has been calculated in \cite{Brambilla:1999qa,Brambilla:1999xf}.

\begin{figure}[htb]
\makebox[0cm]{\phantom b}
\epsfxsize=12truecm \epsfbox{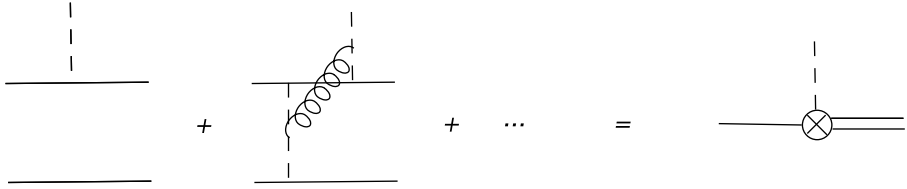}
\caption{Matching condition for $V_A$.
\label{VAmatch}}
\end{figure}

In order to determine the matching coefficients $V^{(0)}_o$ and $V_A$ that enter 
in Eq. (\ref{matchingVs}) besides $V_{s}^{(0)}$, we need two further matching 
conditions. The static octet potential $V^{(0)}_o$ has been 
calculated up to two loops by matching it to a static Wilson loop with 
color matrices in the initial and final states \cite{Kniehl:2004rk}. This gives 
rise to a matching condition similar to Eq. (\ref{matchingVs}) \cite{Brambilla:1999xf}:
\be
 \hbox{$
\displaystyle\lim_{ T\to\infty}\frac{i}{T } 
\ln  \put(0,-5){\epsfxsize=1.2cm \epsffile{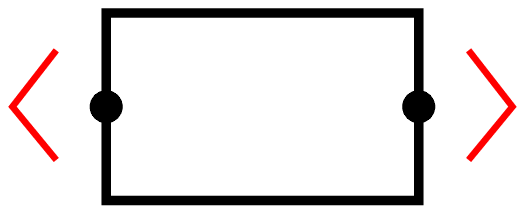}}
\hspace{1.2cm}
=  V_{o}^{(0)}(r,\mu) + \dots \; .$}
\ee
The matching for $V_A$ is described in Fig. \ref{VAmatch}; it gives 
\be
V_A( r,\mu) = 1 + O( \als^2).
\ee

The last ingredient needed in order to calculate Eq.  (\ref{matchingVs}) is the chromoelectric 
correlator $ \langle  {\rm Tr}( {\bf E}(t) \cdot  {\bf E}(0))  \rangle$, where Wilson lines 
connecting the chromoelectric fields are understood. This has been calculated at order $\als$ in 
\cite{Eidemuller:1997bb}.

Since the static Wilson loop is fully known at two loops, the matching condition 
(\ref{matchingVs}) provides the static singlet potential at two loops. 
Moreover, since the static Wilson loop is independent of $\mu$, the right hand-side
of Eq.  (\ref{matchingVs}) should also be $\mu$-independent. 
Therefore, the logarithmic dependence of the 
static potential may be extracted by noting that the $\ln r\mu$, $\ln^2 r\mu$, ...
terms in $V_{s}^{(0)}$ have to cancel against the $\ln (V^{(0)}_o-V^{(0)}_s)/\mu$, 
$\ln^2 (V^{(0)}_o-V^{(0)}_s)/\mu$, ...  $\ln r\mu$, $\ln^2 r\mu$, ... terms in 
$\displaystyle \int_0^\infty \!\!\! dt \, e^{-i t  (V^{(0)}_o-V^{(0)}_s)} \,   
 \langle  {\rm Tr}( {\bf E}(t) \cdot  {\bf E}(0))  \rangle$. 
This leads to a great simplification in the calculation of the logarithmic dependence 
of the static potential: the logarithmic contribution at N$^3$LO 
and the single logarithmic contribution at N$^4$LO may be extracted respectively 
from a one-loop and two-loop  calculation in the EFT.
Finally, we note that the solutions of the renormalization group equations allow the calculation and 
resummation of all logarithmic contributions of a given type (e.g. leading logarithms 
of the type $\als^3 \times (\als\ln \mu r)^n$, next-to-leading logarithms of the type 
$\als^4 \times (\als\ln \mu r)^n$ and so on).

The presently most accurately known fixed-order expression 
of the static singlet potential is 
\bea
 V^{(0)}_s( r,\mu) \!\!
&=& \!\!
- C_F\frac{ \als ( 1/r)}{r} 
\left\{1 + \frac{\als(1/r)}{4\pi}\left[a_1+2\gamma_E\beta_0\right]
\right.
\nn
\\
&& \hspace{-20mm}
+ \left(\frac{\als( 1/r)}{4\pi}\right)^2 
\left[ a_2+ \left(\frac{\pi^2}{3}+4\gamma_E^2\right)\beta_0^2
+2\gamma_E\left(2a_1\beta_0+\beta_1\right)\right]
\nn
\\
&& \hspace{-20mm}
+ 
\left(\frac{\als( 1/r)}{4\pi}\right)^3 
\left[\frac{16\,\pi^2}{3} C_A^3 \,  \ln {r\mu}   + \tilde{a}_3\right] 
\nn
\\
&& \hspace{-20mm}
\left.
+ 
\left(\frac{\als( 1/r)}{4\pi}\right)^4 
\left[  a_{4}^{L2} \ln^2 {r \mu} 
         +  \left( a_{4}^{L}  
         +  \frac{16}{9}\pi^2 \, C_A^3\beta_0 (- 5 + 6 \ln 2)\right)  \ln
	 {r\mu}  + \tilde{a}_{4} \right]
\right\},
\label{Vsn4lo}
\eea
where $C_F= T_F (N_c^2-1)/N_c$, $C_A=N_c$, $T_F=1/2$, $\beta_0 = 11 C_A/3 - 4 T_F n_f/3$, 
$\beta_1 = 34 C_A^2/3$ $ - 20 C_A T_F n_f/3 - 4 C_F T_F n_f$,
$n_f$ is the number of (massless) flavors, $\gamma_E$ is the Euler constant
and $\als$ is the strong coupling constant in the $\MS$ scheme. 
The coefficients $a_1$, $a_2$, $a_4^{L2}$ and $a_4^L$ stand for 
\bea
a_1&=&\frac{31}{9}C_A-\frac{20}{9}T_Fn_f, 
\\
a_2&=& \left(\frac{4343}{162}+4\pi^2-\frac{\pi^4}{4}+\frac{22}{3}\zeta(3)\right)C_A^2
-\left(\frac{1798}{81}+\frac{56}{3}\zeta(3)\right)C_AT_Fn_f
\nn\\
&& -\left(\frac{55}{3}-16\zeta(3)\right)C_FT_Fn_f +\left(\frac{20}{9}T_Fn_f\right)^2,
\\
a_4^{L2}  &=& - \frac{16\pi^2}{3}C_A^3\,\beta_0, 
\\
a_4^L  &=&  16\pi^2C_A^3\left[a_1+2\gamma_E\beta_0 
+ T_F n_f \left( -\frac{40}{27} + \frac{8}{9} \ln 2\right)
\right.
\nn\\
&&\qquad\qquad\qquad\qquad\qquad\qquad\qquad
\left.
+ C_A\left(\frac{149}{27}-\frac{22}{9}\ln 2+\frac{4}{9}\pi^2\right)\right].
\eea
The coefficient $a_1$ was calculated in \cite{Billoire:1979ih}, 
the coefficient $a_2$ in \cite{Peter:1996ig,Peter:1997me,Schroder:1998vy,Kniehl:2001ju}, 
the term proportional to  $\als^4/r \times \ln r\mu$ in \cite{Brambilla:1999qa,Brambilla:1999xf}, 
the coefficient $a_4^{L2}$ in \cite{Pineda:2000gz,Brambilla:2006wp}
and the coefficient $a_4^L$ in \cite{Brambilla:2006wp}. 
The coefficients $\tilde{a}_3$ and $\tilde{a}_4$ are only partially known (see \cite{Brambilla:2006wp}
for discussion and references). The leading logarithmic contributions have been 
resummed to all orders in \cite{Pineda:2000gz}.

Expression (\ref{Vsn4lo}) shows explicitly the Wilson coefficient nature 
of the static potential. It shows a scale dependence, which comes from the renormalization, and 
it satisfies renormalization group equations, which allow to resum potentially large 
$\ln {r\mu}$ terms. Also large contributions of the renormalon type may be analyzed 
in the EFT framework.

By summing Eq. (\ref{Vsn4lo}) to the ultrasoft contributions we get back the static 
Wilson loop, i.e. the energy between two static sources in QCD. This reads
\bea
 E_0(r)  &=&-\frac{C_F\als(1/r)}{r}\Bigg\{1+\frac{\als(1/r)}{4\pi}\left[a_1+2\gamma_E\beta_0\right]
\nn\\
&& +\left(\frac{\als(1/r)}{4\pi}\right)^2
\left[ a_2+ \left(\frac{\pi^2}{3}+4\gamma_E^2\right)\beta_0^2
+2\gamma_E\left(2a_1\beta_0+\beta_1\right)\right]
\nn\\
&&
+\left(\frac{\als(1/r)}{4\pi}\right)^3\left[
   \frac{16\pi^2}{3}C_A^3  \ln \frac{C_A \als(1/r)}{2}  + \tilde{a}^\prime_3\right]
\nn\\
&&
+\left(\frac{\als(1/r)}{4\pi}\right)^4
\left[ a_{4}^{L2}  \ln^2 \frac{C_A \als(1/r)}{2}  
      +a_{4}^{L}   \ln \frac{C_A \als(1/r)}{2}  + \tilde{a}^\prime_{4} \right] \Bigg\}.
\eea
This quantity may be compared with the short-distance behaviour of the static Wilson loop 
provided by lattice calculations, see for instance \cite{Necco:2001xg,Pineda:2002se}.

\section{The Non-perturbative Potential}
\begin{figure}[htb]
\makebox[-6cm]{\phantom b}
\put(60,0){\epsfxsize=6truecm \epsfbox{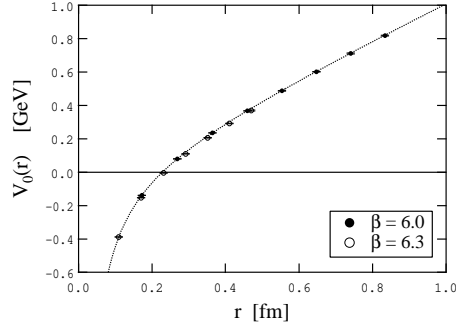}}
\caption{Lattice determination of the right-hand side of 
Eq. (\ref{V0long}), from \cite{Koma:2006fw}.
\label{V0lat}}
\end{figure}

At large distances, $1/r \sim \lQ$, due to confinement, ultrasoft effective degrees of freedom may only 
be colorless objects. If Goldstone bosons are neglected, the color-singlet quarkonium field $S$ 
turns out to be the only dynamical degree of freedom at scales lower than $\lQ$ \cite{Brambilla:1999xf}. The static singlet potential is then simply given by
\be
\hspace{-8mm}
 V_s^{(0)} 
=    \displaystyle\lim_{T\to\infty}\frac{i}{T}
\ln \put(0,-5){\epsfxsize=1.2cm \epsffile{v0.eps}}
\hspace{1.2cm}.
\label{V0long}
\ee
A recent lattice determination is shown in Fig. \ref{V0lat}.

\begin{figure}[htb]
\makebox[-6cm]{\phantom b}
\put(60,0){\epsfxsize=6truecm \epsfbox{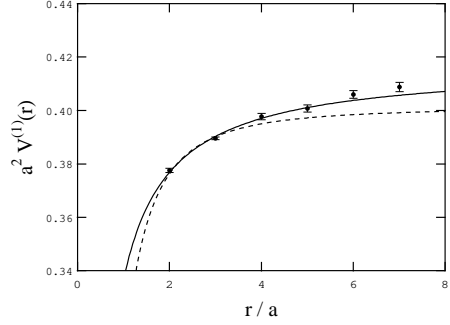}}
\caption{Lattice determination of the right-hand side of Eq. (\ref{V1long}), from \cite{Koma:2006si}.
\label{V1lat}}
\end{figure}

Recently, and for the first time, the leading relativistic correction to the static 
potential has been calculated on the lattice. The existence of a 
possibly large non-perturbative $1/m$ potential, $V_s^{(1)}$, 
was first pointed out in \cite{Brambilla:2000gk}. $V_s^{(1)}$  may be written as a  
static Wilson loop with two chromoelectric field insertions on the same quark line:
\be
\frac{V_s^{(1)}}{m}  = 
- \frac{1}{2 m}\int_0^\infty \!\! dt \, t 
\put(2,-6){\epsfxsize=1.5cm \epsfbox{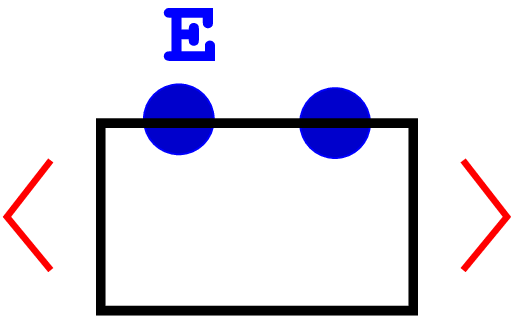}}
\hspace{1.6cm}.
\label{V1long}
\ee
The corresponding lattice determination is shown in Fig. \ref{V1lat}.

Note that, in accordance to power counting arguments, in the long-range, 
the $1/m$ potential may be as large as the static potential and contribute 
with it to the leading-order potential \cite{Brambilla:2000gk}.

\begin{figure}[ht]
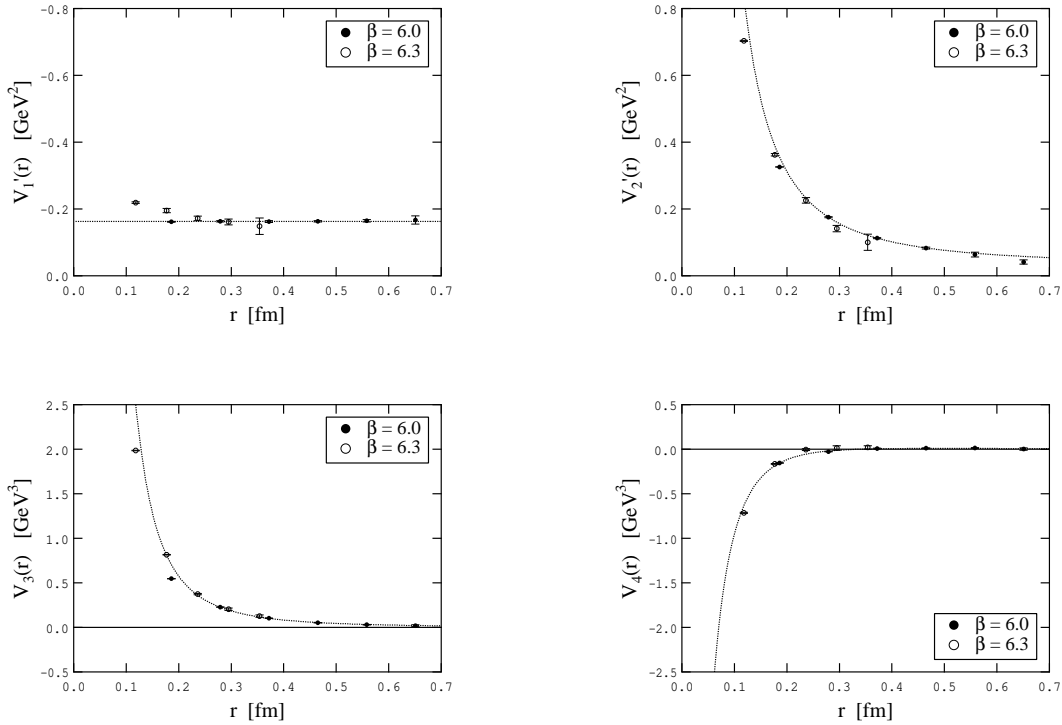

\makebox[-8cm]{\phantom b}
\put(10,0){\epsfxsize=6truecm\epsffile{fig8.EPSF}}
\put(240,0){\epsfxsize=6truecm\epsffile{fig9.EPSF}}
\put(10,-150){\epsfxsize=6truecm\epsffile{fig12.EPSF}}
\put(240,-150){\epsfxsize=6truecm\epsffile{fig13.EPSF}}
\caption{Lattice determination of the spin-dependent $1/m^2$ potentials, from \cite{Koma:2006fw}.
The potentials $V_1^\prime$ and $V_2^\prime$ are spin-orbit potentials, 
the potential $V_3$ is the tensor potential and the potential $V_4$ the spin-spin potential.
\label{V2lat}}
\end{figure}

By the same collaboration, spin-dependent $1/m^2$ potentials have been calculated on the lattice 
with unprecedented precision. Expressions for the spin-dependent potentials in terms 
of static Wilson loops and field-strength insertions have been derived in \cite{Eichten:1980mw,Pineda:2000sz}. 
These have been used to obtain the lattice results shown in Fig. \ref{V2lat}.

In the long range, the spin-orbit potentials show, for the first 
time, deviations from the flux-tube picture of chromoelectric confinement.
Since a fully consistent renormalization of the EFT operators is still missing 
in the lattice analysis, it may be premature to draw any definitive 
conclusion. However, progress has been made recently in this direction.
In \cite{Guazzini:2007bu}, the non-perturbative  renormalization 
of the chromomagnetic operator in the Heavy Quark Effective Theory, 
which crucially enters in all spin-dependent potentials,
has been performed for the first time. A proper operator renormalization is also crucial  
in order to verify an exact relation among the spin-dependent potentials 
required by Lorentz invariance \cite{Gromes:1984ma,Brambilla:2003nt}, which 
was checked in \cite{Koma:2006fw} at the few percent level.

\vspace{-1mm}

\section{Conclusions}
Non-relativistic EFTs provide a rigorous definition of the 
potential between a heavy quark and antiquark (see \cite{Brambilla:2005yk} for systems 
made by two or three heavy quarks).
In the perturbative regime, the potential is a key ingredient for
precision calculations of several threshold observables. 
In the non-perturbative regime, it can be calculated on
the lattice; the corresponding EFT, pNRQCD, may provide lattice studies 
with an alternative to more traditional EFTs with heavy quarks, like NRQCD.

\vspace{-1mm}

\begin{theacknowledgments}
The author acknowledges the financial support obtained inside the Italian 
MIUR program  ``incentivazione alla mobilit\`a di studiosi stranieri e 
italiani residenti all'estero'' and by the European Commission MRTN 
FLAVIA{\it net}  [MRTN-CT-2006-035482].
\end{theacknowledgments}


\vspace{-1mm}

\begin{thebibliography}{99}
\bibitem{Brambilla:2004wf}
  N.~Brambilla {\it et al.},
  {\it Heavy quarkonium physics},
  CERN-2005-005
  [ar\-Xiv:hep-ph/0412158].

\bibitem{Brambilla:2004jw}
  N.~Brambilla, A.~Pineda, J.~Soto and A.~Vairo,
  Rev.\ Mod.\ Phys.\  {\bf 77} (2005) 1423.

\bibitem{Peter:1996ig}
  M.~Peter,
  Phys.\ Rev.\ Lett.\  {\bf 78} (1997) 602.

\bibitem{Peter:1997me}
  M.~Peter,
  Nucl.\ Phys.\ B {\bf 501} (1997) 471.

\bibitem{Schroder:1998vy}
  Y.~Schr\"oder,
  Phys.\ Lett.\ B {\bf 447} (1999) 321.

\bibitem{Kniehl:2001ju}
  B.~A.~Kniehl, A.~A.~Penin, M.~Steinhauser and V.~A.~Smirnov,
  Phys.\ Rev.\ D {\bf 65} (2002) 091503.

\bibitem{Brambilla:1999qa}
  N.~Brambilla, A.~Pineda, J.~Soto and A.~Vairo,
  Phys.\ Rev.\ D {\bf 60} (1999) 091502.

\bibitem{Brambilla:1999xf}
  N.~Brambilla, A.~Pineda, J.~Soto and A.~Vairo,
  Nucl.\ Phys.\ B {\bf 566} (2000) 275.

\bibitem{Kniehl:2004rk}
  B.~A.~Kniehl {\it et al}, 
  Phys.\ Lett.\ B {\bf 607} (2005) 96.

\bibitem{Eidemuller:1997bb}
  M.~Eidem\"uller and M.~Jamin,
  Phys.\ Lett.\ B {\bf 416} (1998) 415.

\bibitem{Billoire:1979ih}
  A.~Billoire,
  Phys.\ Lett.\ B {\bf 92} (1980) 343.

\bibitem{Pineda:2000gz}
  A.~Pineda and J.~Soto,
  Phys.\ Lett.\ B {\bf 495} (2000) 323.

\bibitem{Brambilla:2006wp}
  N.~Brambilla, X.~Garcia i Tormo, J.~Soto and A.~Vairo,
  Phys.\ Lett.\  B {\bf 647} (2007) 185.

\bibitem{Necco:2001xg}
  S.~Necco and R.~Sommer,
  Nucl.\ Phys.\ B {\bf 622} (2002) 328.

\bibitem{Pineda:2002se}
  A.~Pineda,
  J.\ Phys.\ G {\bf 29} (2003) 371.

\bibitem{Koma:2006fw}
  Y.~Koma and M.~Koma,
  Nucl.\ Phys.\  B {\bf 769} (2007) 79.

\bibitem{Brambilla:2000gk}
  N.~Brambilla, A.~Pineda, J.~Soto and A.~Vairo,
  Phys.\ Rev.\  D {\bf 63} (2001) 014023.

\bibitem{Koma:2006si}
  Y.~Koma, M.~Koma and H.~Wittig,
  Phys.\ Rev.\ Lett.\  {\bf 97} (2006) 122003. 

\bibitem{Eichten:1980mw}
  E.~Eichten and F.~Feinberg,
  Phys.\ Rev.\  D {\bf 23} (1981) 2724. 

\bibitem{Pineda:2000sz}
  A.~Pineda and A.~Vairo,
  Phys.\ Rev.\  D {\bf 63} (2001) 054007 
  [Erratum-ibid.\  D {\bf 64} (2001) 039902].

\bibitem{Guazzini:2007bu}
  D.~Guazzini, H.~B.~Meyer and R.~Sommer  [ALPHA Collaboration],
  arXiv:0705.1809 [hep-lat].

\bibitem{Gromes:1984ma}
  D.~Gromes,
  Z.\ Phys.\  C {\bf 26} (1984) 401. 

\bibitem{Brambilla:2003nt}
  N.~Brambilla, D.~Gromes and A.~Vairo,
  Phys.\ Lett.\ B {\bf 576} (2003) 314.

\bibitem{Brambilla:2005yk}
  N.~Brambilla, A.~Vairo and T.~R\"osch,
  Phys.\ Rev.\  D {\bf 72} (2005) 034021.

\end{thebibliography}
\end{document}